\documentclass[11pt,twoside]{article}

\usepackage{asp2004}
\usepackage{epsf}
\usepackage{psfig}
\usepackage{lscape}
\usepackage{multicol}

\begin{document}
\title{Reionization on ice}   
\author{C. C. Dudley,  M. Imanishi \& P. R. Maloney}   
\affil{Naval Research Laboratory; NAO, Japan; CASA, U. of Colorado}    

\begin{abstract} 
The case for substantial far infrared ice emission in local
ultraluminous infrared galaxies, expected based on the
presence of mid-infrared ice absorption in their spectra and
the known far infrared optical properties of ice, is still
largely unsupported by direct observation owing to
insufficient far infrared spectral coverage.  Some marginal
supportive evidence is presented here.

A clear consequence of far infrared ice emission is the need
to extend the range of redshifts considered for
submillimeter sources.  This is demonstrated via the example
of HDF 850.1.

The solid phase of the ISM during reionization may be
dominated by ice, and this could lead to the presence of
reionization sources in submillimeter source catalogs.
Submillimeter sources not detected at 24 $\mu$m in the
GOODS-N field are examined.  Two candidate reionization
sources are identified at 3.6 $\mu$m through possible
Gunn-Peterson saturation in the $Z$ band.
\end{abstract}

\section*{}\phantom{.}\vskip-2.5cm
\section{Ice emission}  

The discovery of ice absorption features in the spectra of
ultraluminous infrared galaxies (ULIRGs; Spoon et al. 2002;
Imanishi \& Maloney 2003; Imanishi, Dudley \& Maloney 2006)
implies that ice is emitting in the far infrared (FIR).  The
obervational consequences are determined by the portion of
FIR emission that is owing to mid-infrared absorption (often
the majority) and the ratio of mantle-to-core volumes of the
emitting dust as well as the FIR optical depth.  For ULIRG
FIR dust temperatures one expects the broad 150 $\mu$m phonon
mode (Berti et al. 1969; Curtis et al. 2005) of ice to modify the
FIR spectral energy distribution (SED) through emission.  An
example is given in Fig. 1a.

This changes at the higher dust temperatures required at higher
redshift where the CMB temperature exceeds (optically thin)
ULIRG temperatures.  Then, shorter wavelength ice features
are more important to SED modification.  Fig.~1b.~shows HDF
850.1, the brightest submillimeter source in the {\it Hubble}
Deep Field, plotted against an ice emission model.  The
redshift, $z=12.6$, is 13 confidence intervals beyond the
estimate of Dunlop et al. (2004): $z=4.1$.  Thus, ice emission could
greatly broaden the range of redshifts that need to be
considered for submillimeter sources.  Fig.~3a.~shows the
expected 850 $\mu$m brightness of sources with the
luminosity of ULIRG IRAS 00188-0856 if they are similar to Arp 220
(long dashed line), have substantial ice and a temperature
of 30 K (solid line), or are similar to the model shown in
Fig. 1b. (50 K; short dashed line). It is notable that the
last model selects for $z=13$. 
 
\begin{figure}[!ht]
\psfig{figure=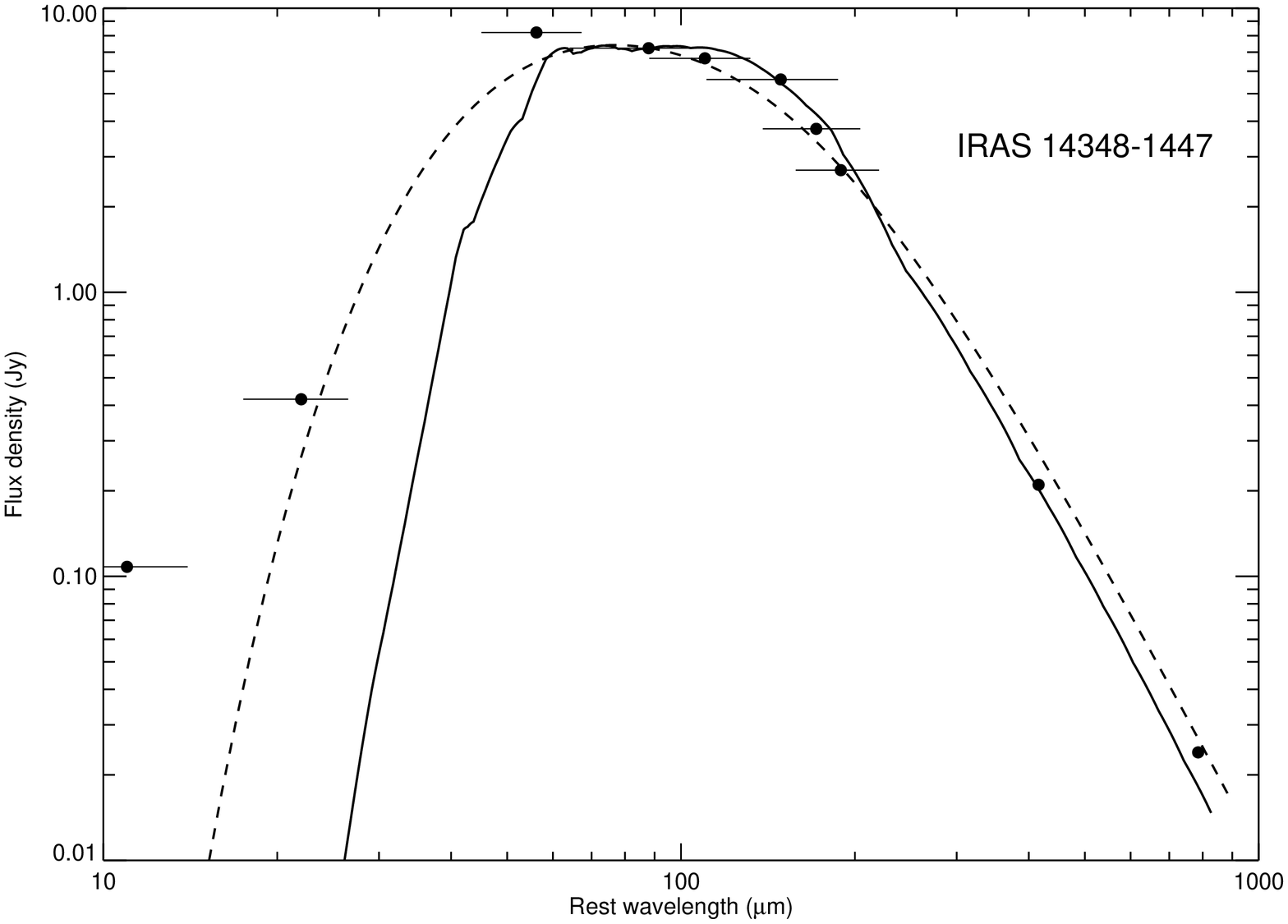,height=4cm}\vskip -4cm\hskip7cm\psfig{figure=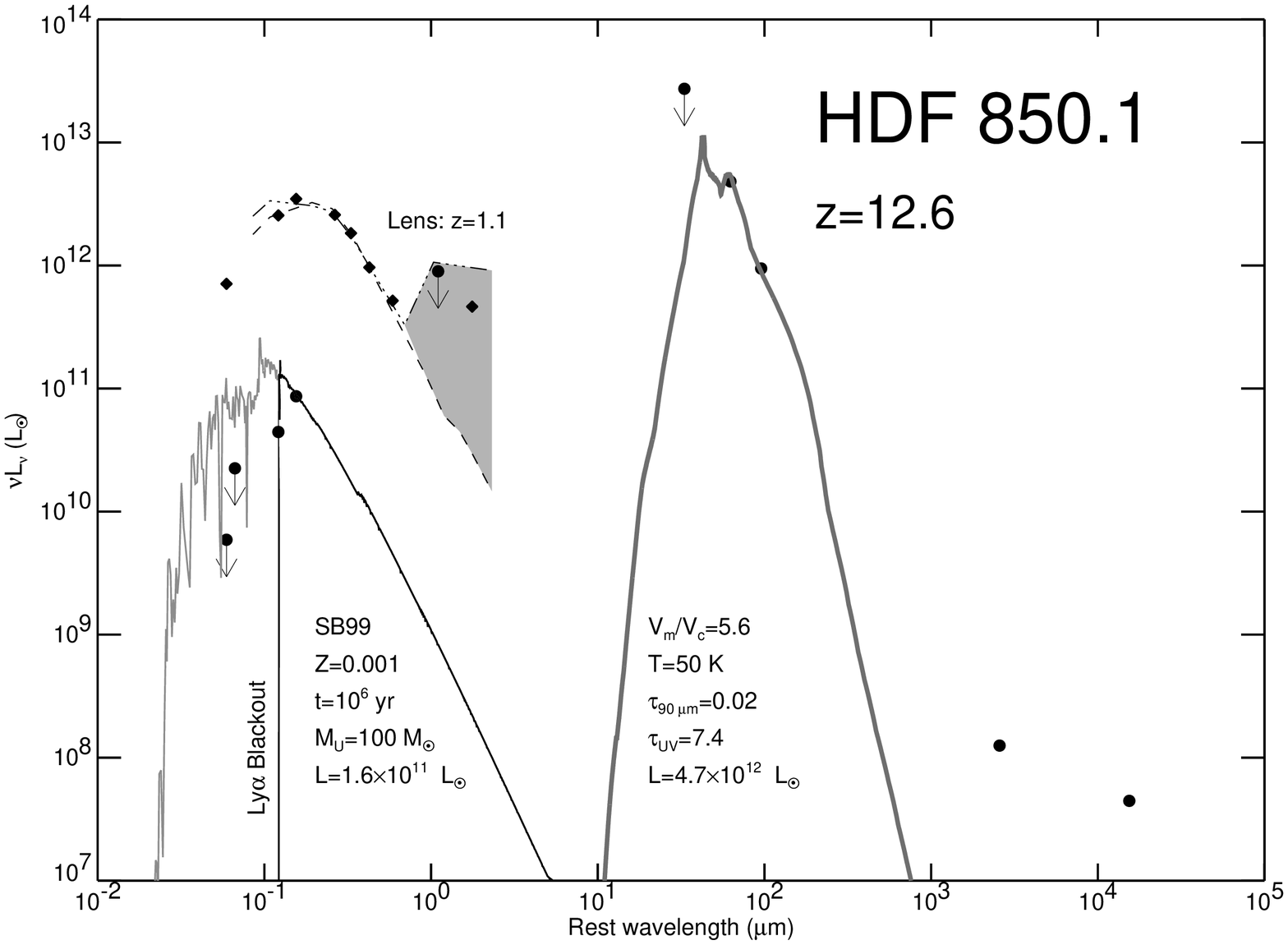,height=4cm}
\vskip0cm\noindent \parbox{6.5cm}{\footnotesize Fig.~1a. Two fits for the
FIR SED of ULIRG IRAS 14348-1447 are shown.  The dashed line
reproduces the fit given by Klaas et al. 2001 and has a 100
$\mu$m optical depth of 5 and $\beta=2$ while the solid line
is an ice mantle-core emission model taken from Aannestad
(1975).  The broad feature centered near 150 $\mu$m is
hinted at in the ISOPHOT data though one fit is not better than the other.  This source shows 6 $\mu$m absorption {\it e.g.} Charmandaris et al. (2002).}\hskip 0.3
cm\parbox{6.5cm}{\footnotesize Fig.~1b. Data from Dunlop et
al. (2004) plus a new $Z$ band limit (Dudley et al. 2006;
filled circles; lensing corrected) are plotted together with
new {\it Spitzer} GOODS data for the lensing galaxy (filled
diamonds; relative scale). Solid lines show an ice emission model and a
starburst model (Leitherer et al. 1999).  Dashed lines and
the shaded region represent evolution of the lensing
eliptical galaxy {\it e.g.} Athey et al. (2002).}
\vskip -0.6cm\end{figure}
\vskip -6cm
\section{What is the early ISM like?}  
When metals are first released into the interstellar
medium the oxygen-to-carbon ratio is high and the $^{56}$Ni production
is frequently low.  For a 186 M$_\odot$ progenitor (90 M$_\odot$ core)
the yield ratios are O:Si:C:S:$^{56}$Ni=123:29:15:10:1 by number
(Heger \& Woosley 2002) so that the onset of
dust formation may be rapid and occur in an oxygen rich
environment. If grain temperature is determined by kinetic and chemical
interactions when the mainly oxygen ejecta mix with the surrounding
mainly hydrogen atmosphere (Baraffe, Heger \& Woosley 2001)
for example, rather than by ongoing radioactivity, grain surface and
gas phase chemistry might lead to the condensation of ice mantles at
temperatures high enough to lead to the formation of crystalline rather
than amorphous ice such as is observed to occur in some evolved stars.

If all or most oxygen enters the solid state, the volume of
ice at this early phase would be larger than the volume of
more refractory material.  Further, if the injection of such
material is needed to facilitate the formation of the
generation of stars that is proposed to be 4 Gyr old at
$z=$1.5 (Jimenez et al. 2000) or $>$2 Gyr old at $z=$2.5
(Stockton et al. 2004) then the formation interval, if set
by this initial enrichment, may be brief enough that a
substantial fraction of stars are found simultaneously
within their natal envelopes which may yet contain
crystalline ice.  

\vskip 0.5cm
\begin{figure}[!ht]
\psfig{figure=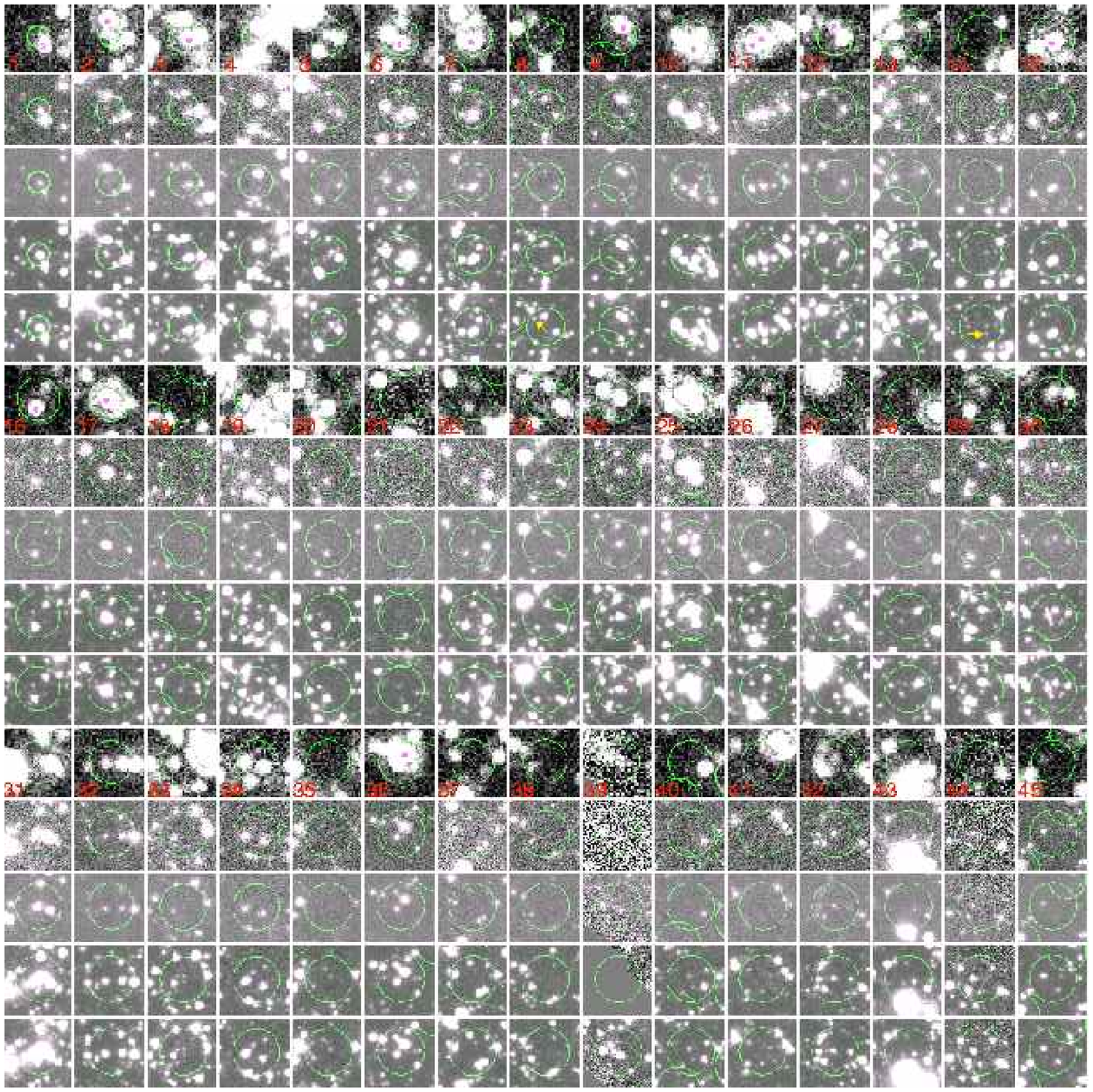,height=8cm,angle=90} 
\noindent \vskip-8cm\hskip8.7cm\parbox{4.5cm}{\footnotesize Fig.~2. {\it Spitzer}
GOODS data for the 45 Wang et al. (2004) HDFGOODS
submillimeter sources.  The darker red-numbered bands are MIPS
24 $\mu$m images, and below these are the corresponding 8,
5.8, 4.5 and 3.6 $\mu$m IRAC data.  The numbers are as
given by Wang et al. with their (green) 2$\sigma$ radii
positional error circles.  Dots (pink) are radio positions
of claimed (Chapman et al. 2005; Wang et al.) associations.
HDFGOODS 850 $\mu$m sources 8, 14, 18, 21, 35 and 45 do not
have obvious 24 $\mu$m couterparts within the error circles.
Yellow arrows indicate sources we have discovered at 3.6 $\mu$m
that do not have $Z$ band counterparts, similar to HDF 850.1}
\vskip-0.3cm
\end{figure}

\vskip -3cm\section{Does SCUBA see to reionization?}

The positional error cicles shown in Fig. 2 for SCUBA 850
$\mu$m sources cataloged by Wang et al. (2004) are
remarkable because so many of them are empty at 24 $\mu$m.
Six are definitely clear of obvious sources and an
additional four (sources 27, 28, 41 and 44) are borderline.
If every SCUBA source has a 24 $\mu$m counterpart, we expect
this situation to occur 2.25 times using 2$\sigma$ radii
error circles.  And, that is in the absence of coincidental
association.   So, some number greater than zero of the
850 $\mu$m sources may not have 24 $\mu$m counterparts.

At the sensitivity of the GOODS MIPS image that implies
$z>4$ using ULIRG templates.  Chapman et al. (2005) argue
that radio non-detected submillimeter sources have a similar
redshift distribution to those radio detected sources for
which they have obtained optical spectra.  However, their
argument depends on the SEDs of the sources being similar to
normal galaxies, and Fig. 3b. shows that this may not be the
case.  So, the MIPS data together with a reanalysis of
sources with known redshifts suggest a higher redshift
population could be present in submillimeter catalogs.

\begin{figure}[ht!]
\vskip-0.3cm\psfig{figure=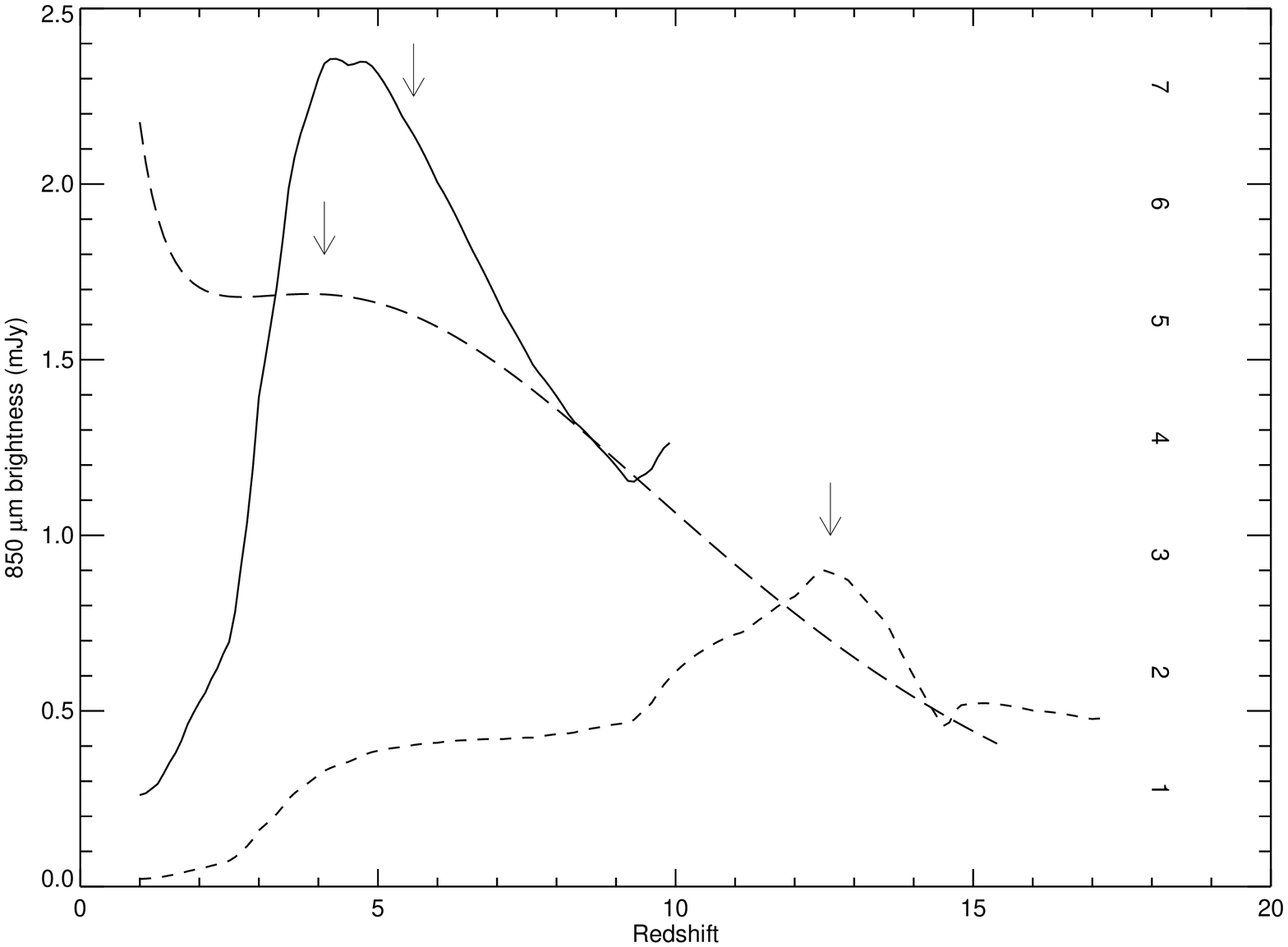,height=4cm}\vskip -4cm\hskip7cm\psfig{figure=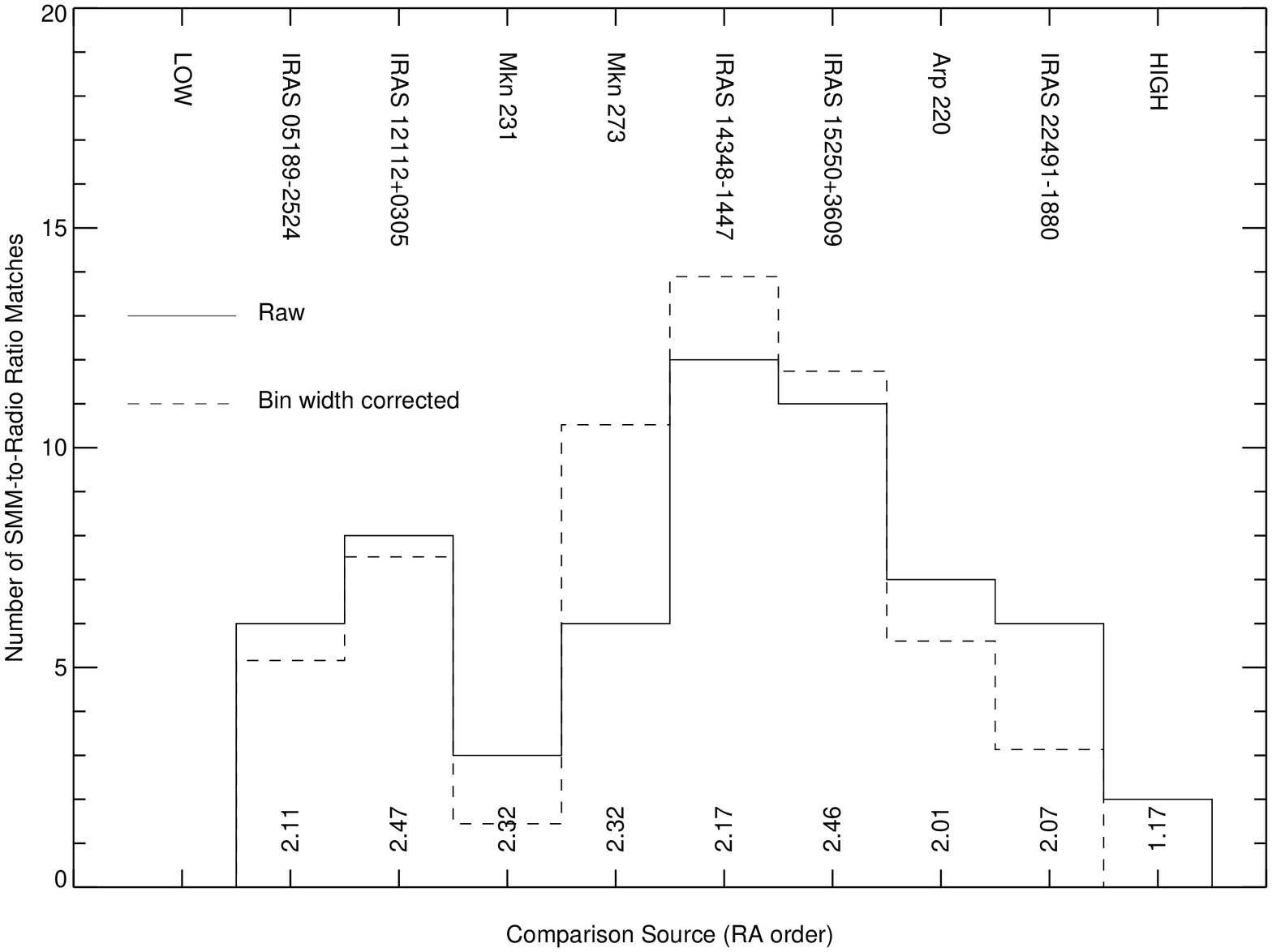,height=4cm,angle=90}
\noindent \parbox{6.5cm}{\footnotesize Fig.~3a. 850 $\mu$m brightness
is plotted against redshift for a typical SED model (long
dashed line; Scoville et al. 1991), an ice emission model for
IRAS 00188-0856 (solid line; Dudley et al. 2006) and the
model shown in Fig. 1b.  The right hand scale accounts for
lensing amplification for HDF 850.1. The arrows are
redshift estimates for this source from Dunlop et al. and Dudley et al.}\hskip 0.3
cm\parbox{6.5cm}{\footnotesize Fig.~3b.~850 $\mu$m-to-radio ratios are
caclulated for each of the listed ULIRGs at the redshifts of the sources
cataloged by Chapman et al. (2005) for $1<z<3.2$.  The
Chapman et al. sources are assigned to bins based on
the ULIRG distribution.  The histogram is shown by the
solid line.  The dashed line shows a bin width correction.  Bin
average redshifts are listed above the lower axis.}
\vskip -0.3cm\end{figure}

We find sources in the 24 $\mu$m blank fields for
HDFGOODS 850-8 and 850-14 which are similar to HDF 850.1 in
that they are detected in the "NIR" (3.6 $\mu$m) but not at
$I$ or $Z$ band (Fig. 4).  $I-[3.6]>5$ and 4.5 (Vega)
respectively.  Should ice play the role suggested in
Fig. 1b. in these sources, making them detectable at 850
$\mu$m but not too luminous to be precursors of the known
old elipticals at $z=$1.5 and 2.5, then these could be
reionization sources.  If Gunn-Peterson saturation explains
the $Z$ band non-detections then the redshifts of these
sources would be greater than 7 and thus larger that the
$z\sim$6 onset of the Gunn-Peterson trough (Fan et al. 2004)
while the redshift of last star formation is $z\sim$13 for
the old ellipticals in a WMAP cosmology (Spergel et al. 2003).
This coincides with the redshift range for reionization estimated
by Kogut et al. (2003).  These sources may prove easier to observe
than HDF 850.1 owing to the absence of a lens.

\vskip 0.5cm
\footnotesize\begin{figure}
\vskip-0.2cm\psfig{figure=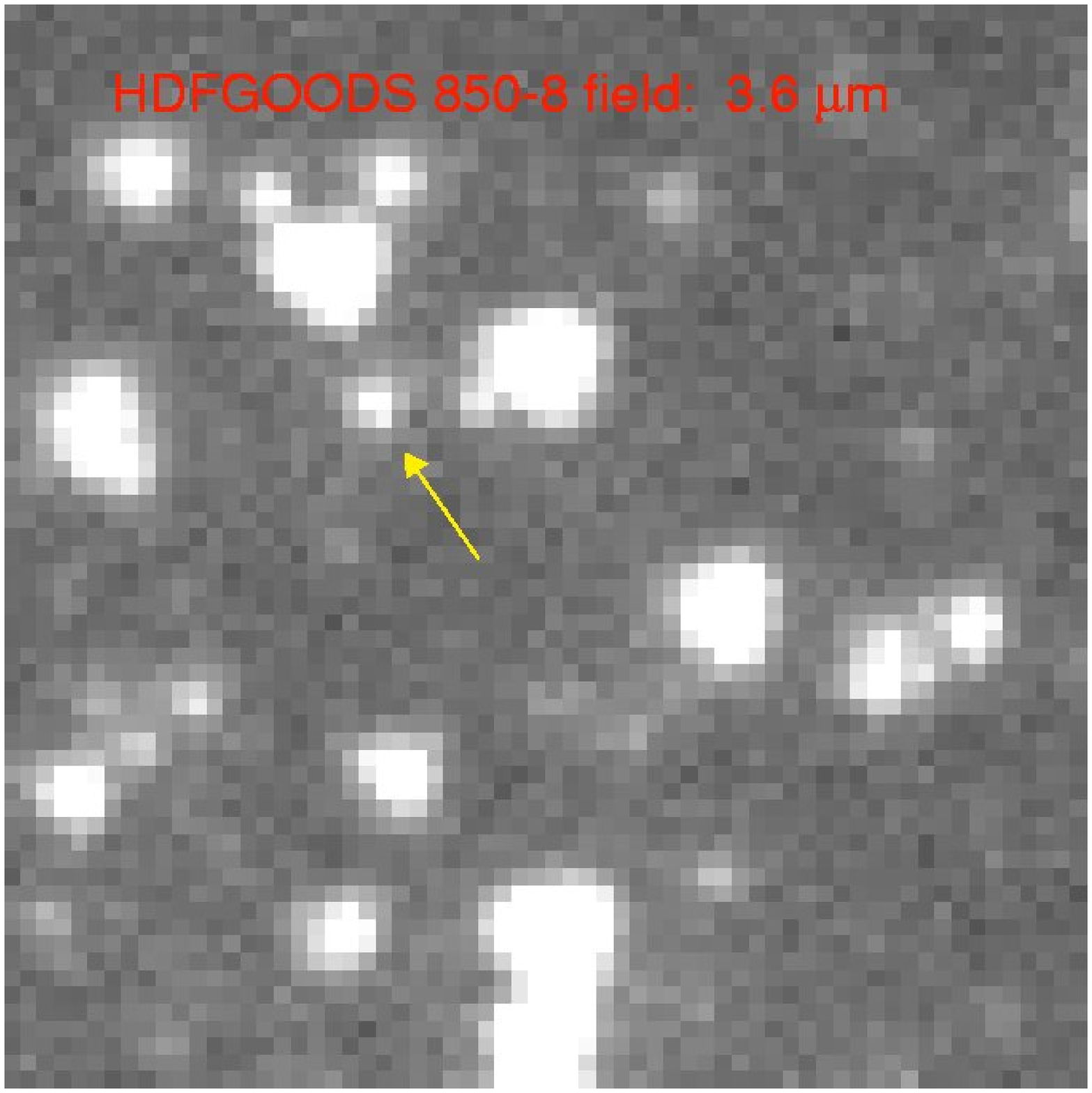,height=4cm,angle=90}\vskip-4cm\hskip 4.3cm\psfig{figure=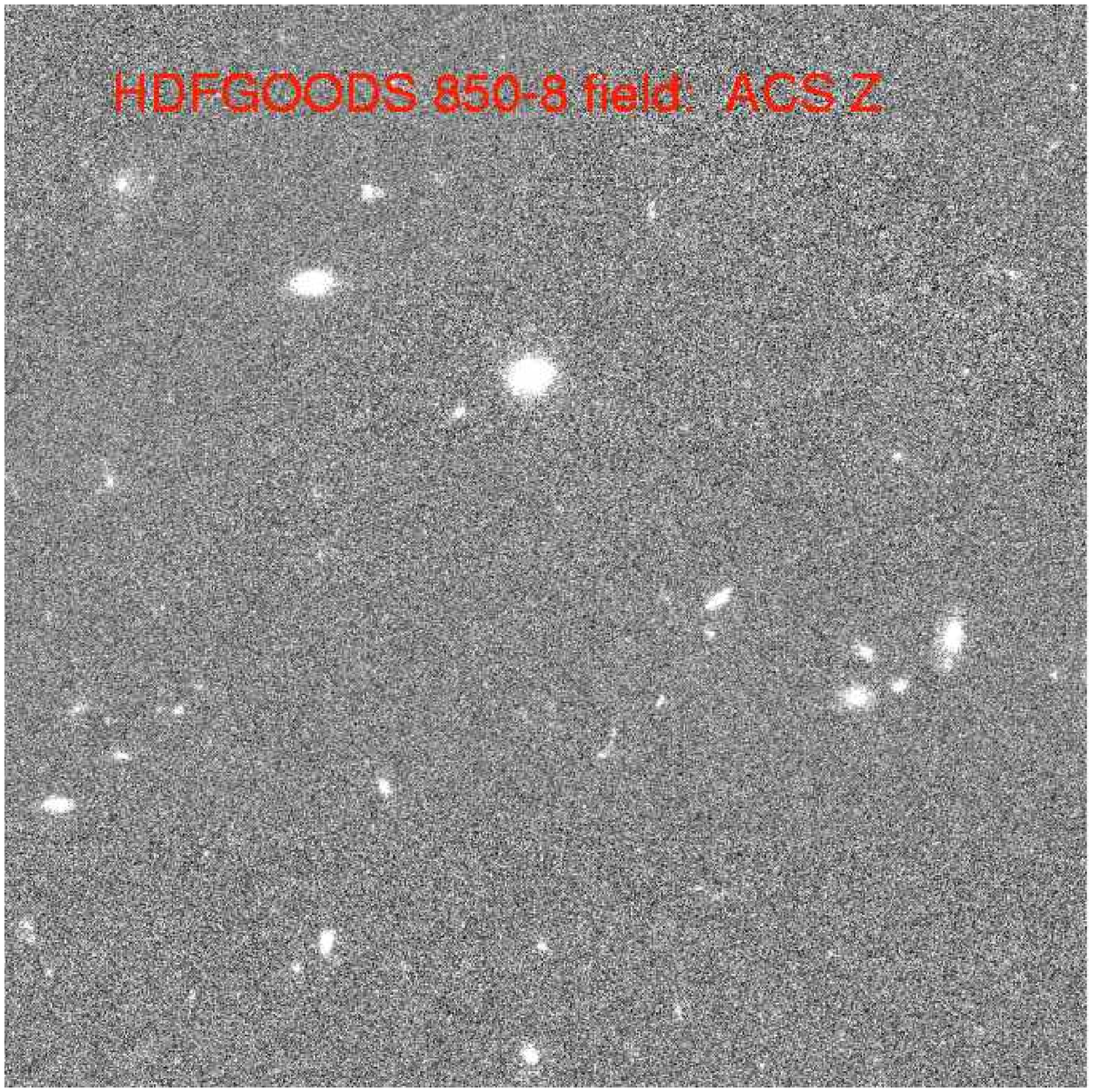,height=4cm,angle=90}\vskip-4cm\hskip 8.6cm\psfig{figure=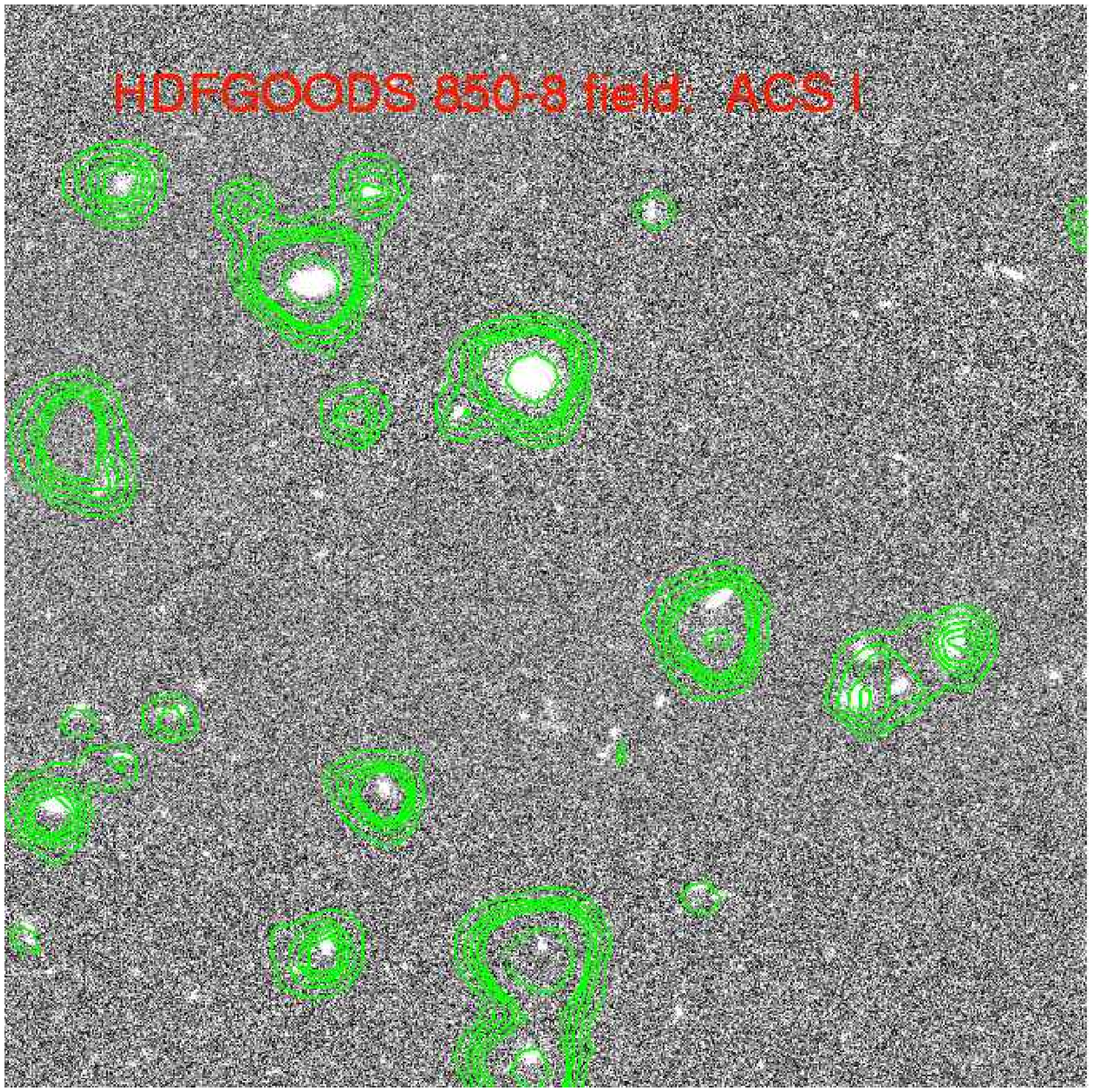,height=4cm,angle=90}\vskip0.3cm\psfig{figure=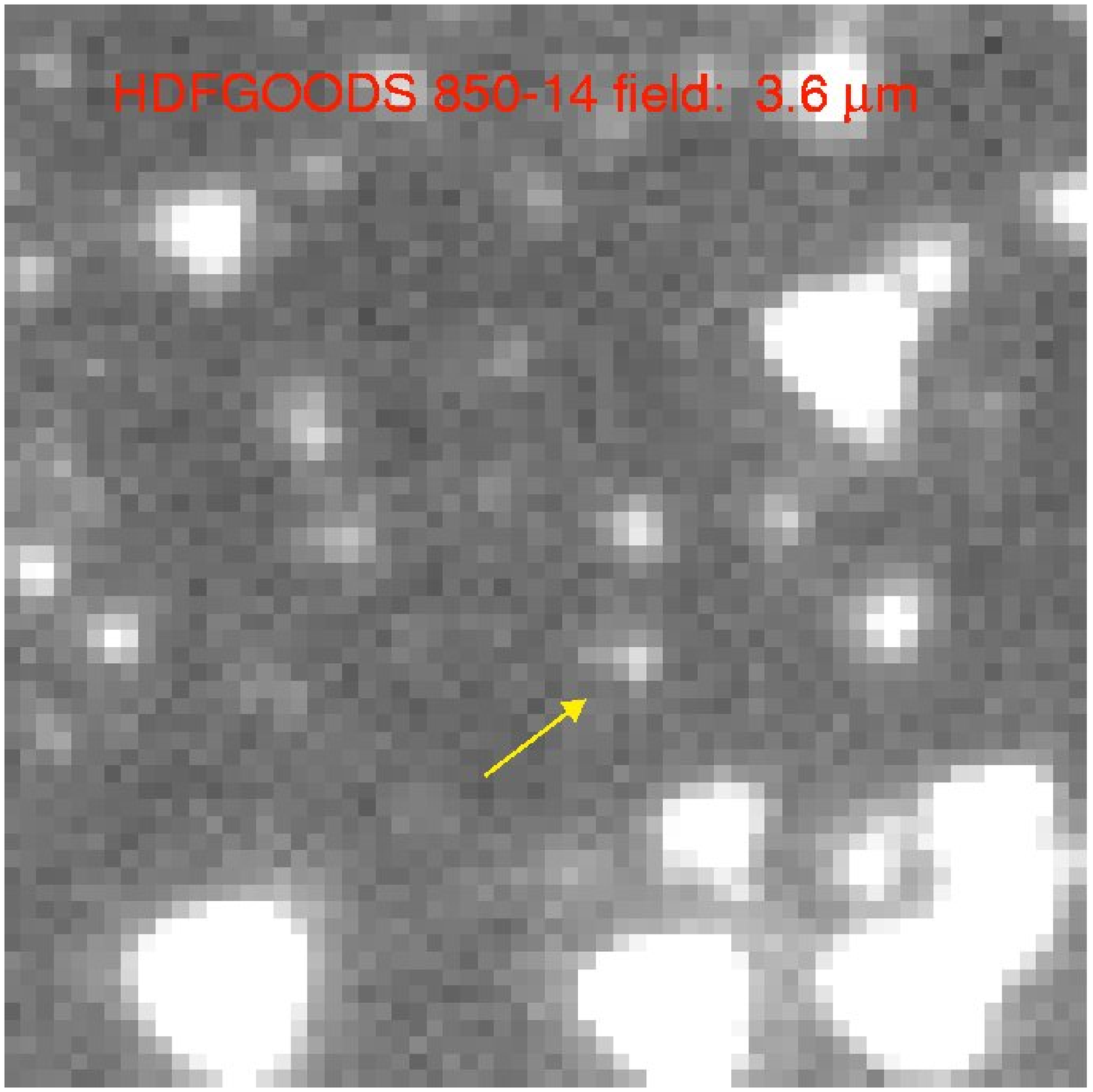,height=4cm,angle=90}\vskip-4cm\hskip 4.3cm\psfig{figure=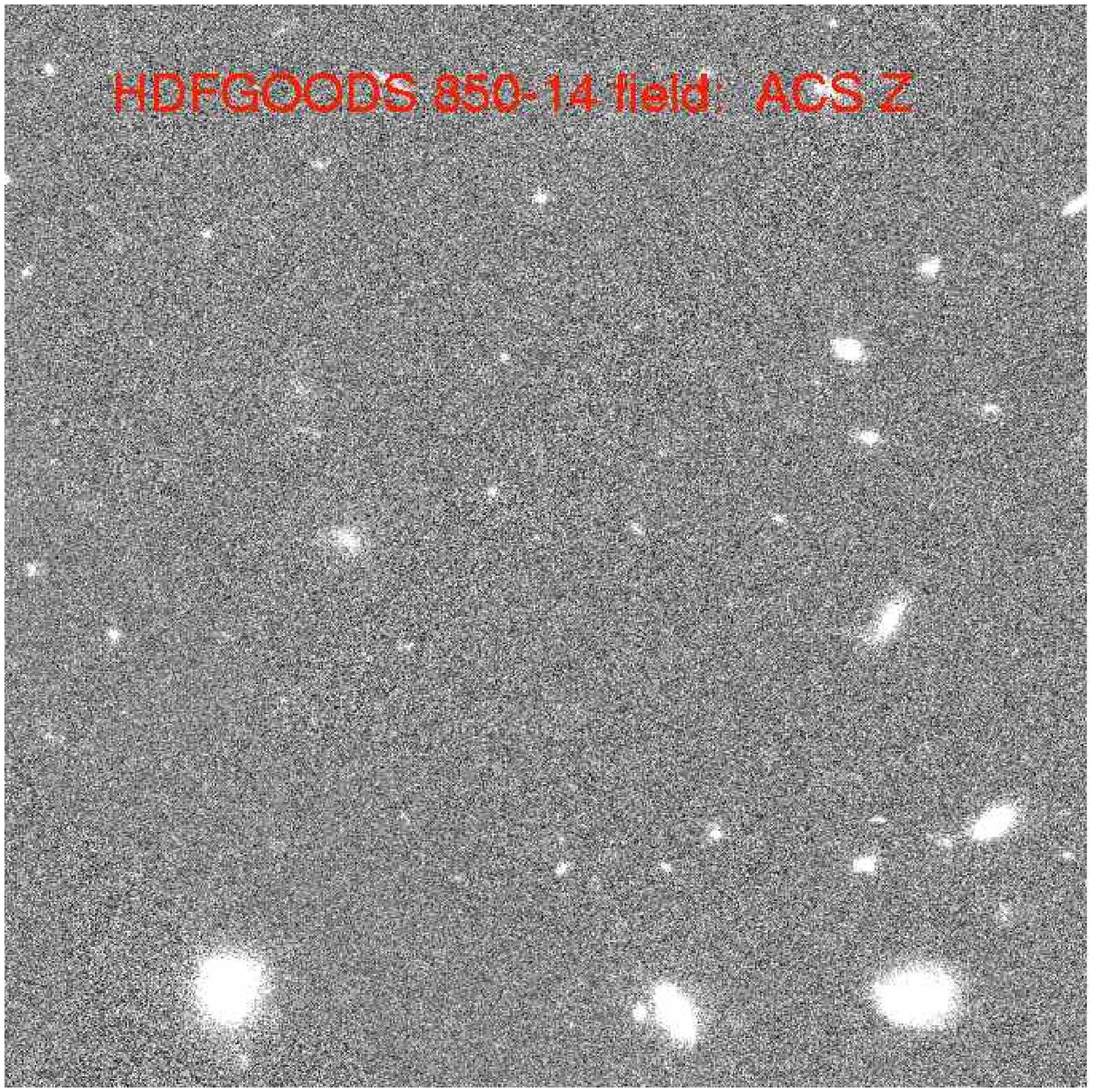,height=4cm,angle=90}\vskip-4cm\hskip 8.6cm\psfig{figure=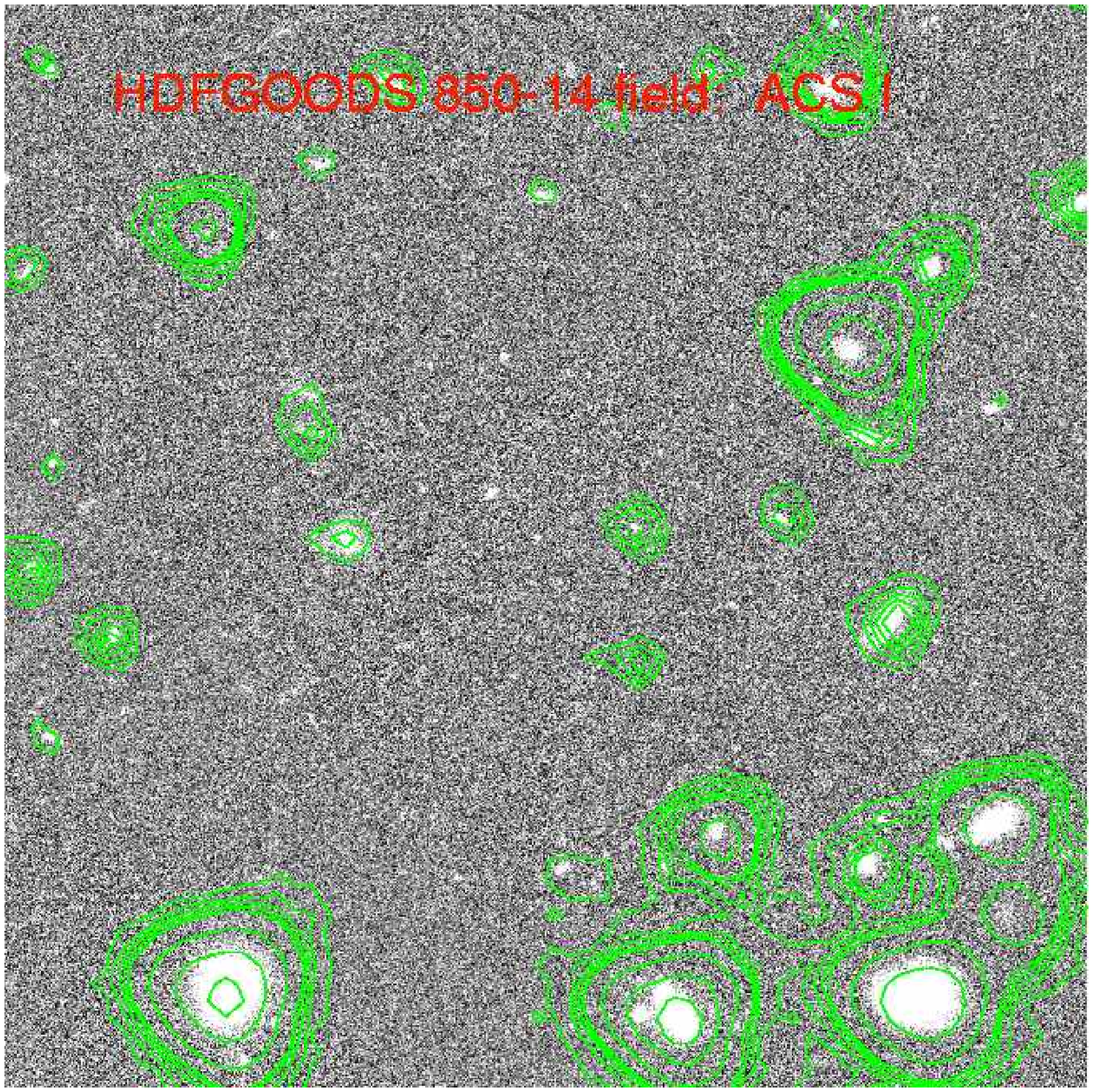,height=4cm,angle=90}
\noindent{\footnotesize Fig.~4. {\it Spitzer} IRAC 3.6 $\mu$m images for the fields of HDFGOODS 850-8 and 850-14 are reproduced from Fig. 2 together with {\it Hubble} ACS $Z$ and $I$ band images.  The 3.6 $\mu$m sources indicated with the yellow arrows are absent at the shorter wavelengths. Green contours are 3.6 $\mu$m data.}
\vskip-0.6cm
\end{figure}\normalsize

\acknowledgements 
Thanks to M. Dickinson (GOODS PI) and D. Scott (however Wang et al. astro-ph/0512347 do not concede) for helpful discussion at the conference.  Infrared Astronomy at the Naval Reseach Laboratory is supported by the Office of Naval Research (USA).
\vskip-0.3cm\footnotesize\begin{multicols}{2}
\noindent{Aannestad, P. A. 1975, ApJ, 200, 30\\ Athey, A. et al., 2002, ApJ, 571, 272 \\ Baraffe, I, Heger, A., Woosley, S. E., 2001, ApJ, 550, 890\\ Bertie, J. E., Labb\'{e}, H. J. \& Whalley, E. 1969, J Chem Phys, 50, 4501\\Chapman, S. C., Blain, A. W., Smail, I., Ivison, R. J., 2005, ApJ, 622, 772\\Charmandaris, V. et al. 2002, A\&A, 391, 429\\Curtis, D. B. et al., 2005, App. Opt., 44, 4102 \\Dudley, C.C., Imanishi, M. Maloney, P.R., 2006, MNRAS submitted\\Dunlop, J. S. et al. 2004, MNRAS, 350, 769\\Fan, X., et al. 2004, AJ, 128, 515\\Heger, A. \& Woosley, S. E. 2002, ApJ, 567, 532\\Imanishi, M., Maloney, P. R., 2003, ApJ, 588, 165\\Imanishi, M., Dudley, C. C., Maloney, P. R., 2006, ApJ, 637, 114\\Jimenez et al. 2000, ApJ, 532, 15\\Klaas, U. et al. 2001, A\&A, 379, 823\\Kogut, A. et al. 2003, ApJS, 148, 161\\Leitherer, C. et al. 1999, ApJS, 123, 3\\Scoville, N. Z. et al. 1991, ApJ, 366, L8\\Spergel, D. N. et al. 2003, ApJS, 148, 175\\Spoon, H. W. W. et al. 2002, A\&A, 414, 873\\Stockton, A., Canalizo, G., Maihara, T. 2004, ApJ, 605, 37\\Wang, W.-H., Cowie, L. L., Barger, A. J. 2004, ApJ, 613, 655} \end{multicols}

\end{document}